\documentclass[aps,prb,showpacs,groupedaddresss,superscriptaddress, twocolumn]{revtex4}
\usepackage{graphicx}
\usepackage{amsmath}
\usepackage{bm}
\begin{document}

\title{Perfect valley filter based on topological phase in disordered $\rm{Sb}$ Monolayer Heterostructure}
\author {Shu-guang Cheng}
\affiliation{Department of Physics, Northwest University, Xi'an 710069, People's Republic of China}
\affiliation{Shaanxi Key Laboratory for Theoretical Physics Frontiers, Xian 710069, People's Republic of China}
\author {Rui-zhi Zhang}
\affiliation{Department of Physics, Northwest University, Xi'an 710069, People's Republic of China}
\author {Jiaojiao Zhou}
\affiliation{College of Physics, Optoelectronics and Energy, Soochow University, Suzhou, 215006, People's Republic of China}
\author {Hua Jiang$^\dagger$}
\email{jianghuaphy@suda.edu.cn}
\affiliation{College of Physics, Optoelectronics and Energy, Soochow University, Suzhou, 215006, People's Republic of China}
\affiliation{Institute for Advanced Study, Soochow University, Suzhou, 215006, People's Republic of China}
\author {Qing-Feng Sun}
\affiliation{International Center for Quantum Materials, School of Physics, Peking University, Beijing 100871, People's Republic of China}
\affiliation{Collaborative Innovation Center of Quantum Matter, Beijing 100871, People's Republic of China}

\begin{abstract}
The hydrogenated $\rm{Sb}$ monolayer epitaxially grown on a $\rm{LaFeO_3}$
substrate is a novel type of two-dimensional material hosting quantum spin-quantum
anomalous Hall (QS-QAH) states. For a device formed by $\rm{Sb}$ monolayer ribbon, the QAH edge states,
belong to a single valley, are located at opposite edges of the ribbon. The QSH edge states, on the other hand, belong to the other valley and are distributed in a very narrow region at the same edge.
In this paper, we find such material can be used to fabricate perfect valley filter.
Adopting scattering matrix method and Green's function method,
the valley resolved transport and spatial distribution of local current are calculated, in the present of Anderson disorder, edge defects and edge deformations.
The numerical results demonstrate that, in the presence of above three types of disorder with moderate strength, the carriers can flow disspationless with nearly perfect valley polarization.
Moreover, when the device becomes longer, the transport current does not decrease
while the valley filter works even better.
The origin is that the disorder can destroy the QSH edge states, but the valley-polarized
QAH edge states can well hold.
%These results are also supported by the results of local current distribution.
\end{abstract}
\pacs{73.63.-b, 72.80.Ng, 73.20.At}
\maketitle

\section{INTRODUCTION}

Two dimensional materials have captured great attention in the past decades
since the discovery of graphene, the single layer hexagonal lattice of carbon atoms.
One peculiar character of hexagonal structure material
(such as graphene,\cite{Q01} $\rm{MoS_2}$,\cite{Q02} silicene\cite{Q03,Q04} etc)
is the valley index in band structure. In analogy to electron's spin,
valley index provides another degree of freedom for quantum information manipulation, named valleytronics.\cite{A11,A06,A05,A10,A12,Q05,Q06,Q07,A09,Q08,Q09,Q10,Q11,Q12,Q13,Q14,Q15,Q16,Q17,Q18,aSong2,A13,A14,A16,Q19,Q20,Q21,Q22,Q23} A valley filter, which generates valley polarized carriers, is the key device for the application of valleytronics.\cite{Q09,Q10,Q11,Q12,Q13} A valley filter can be formed by the way of optical pumping in $\rm{MoS_2}$,\cite{Q05,Q06,Q07,A09,Q08} at the domain wall of materials with reversal inversion asymmetry\cite{Q14,Q15,Q16,Q17,Q18,aSong2,A13,A14,A16} or through valley Hall effect.\cite{Q05,Q19,Q20,Q21,Q22,Q23} Valley Hall effect is predicted when the inversion symmetry is broken in monolayer graphene or few-layer graphene. Recently in experiment the valley Hall effect is reported in bilayer graphene with nonlocal measurement.\cite{Q22,Q23}

One of the most important indexes that characterize a valley filter is the valley current polarization.\cite{Q12,Q13,A17,A18} A perfect valley filter requires the following two key points:
i) it should be a material experimentally accessible and, ii) nearly perfect valley polarization and large magnitude of the current are asked as well. What's more, the transport of valley polarized current should be dissipationless, thus is robust against various kinds of disorders and edge deformations. In conventional valley materials, the inevitable disorders introduce intervalley scattering and backscattering.\cite{Q12,Q13,A01,A02,A07,A08,A19} Thus, the performance of the valley filters could be bad, especially in longer valley filters. Such phenomenon are observed in the experiments of graphene\cite{A01,A02} and is studied by some of the present authors as well.\cite{Q13} To avoid the disorder induced backscattering, the topological protected states may be better candidates for building perfect valley filters.

Using a toy model, Pan \textit{et al.} proposed the valley-polarized quantum anomalous Hall (QAH) states in silicene.\cite{Q24} Under proper disorder, two pairs of edge states are destroyed through intervalley scattering, leaving the QAH edge states in a single valley.\cite{Q10}
However, to achieve such goal, the parameters required are unrealistic experimentally. Furthermore, it is also predicted that a valley filter exists at the boundary between a QAH insulator and a quantum valley Hall insulator based on hexagonal lattice models.\cite{Q11} It is rather difficult to generate two different topological states in a single sample. In above two proposals, the perfect valley states are more like theoretical hypothesis other than realistic proposals. Moreover, the direct investigation of how to manipulate the valley transport in these two system still remains to be investigated.

Soon after the fabrication of monolayer $\rm{Sb}$ in experiment,\cite{A03, A04} quantum spin-quantum anomalous Hall (QS-QAH)\cite{Q26} is predicted in hydrogenated $\rm{Sb}$ monolayer based heterostructure using \textit{ab initio} calculation.\cite{Q27}
In such model, hydrogenated $\rm{Sb}$ monolayer is epitaxially grown on a $\rm{LaFeO_3}$ substrate, in which the spin-orbit coupling of $p_x$ and $p_y$ is strong. The combination of staggering exchange field and spin-orbit coupling leads to a nontrivial topological phase with large bulk gap.
Considering the spatial distribution of the edge states (see Fig. \ref{Fig1} b), one speculate that the valley polarized QAH states are robust and the QSH states are fragile in the presence disorder, which can be used as valley filters.

In this paper, we present the pure valley polarized QAH effect in a $\rm{Sb}$ monolayer ribbon,
in the presence of Anderson disorder, edge defects, or edge deformations.
Using a tight-binding model, the scattering matrix method and Green's function methods are applied to calculate the valley resolved transport coefficients and local current distributions.
In a clean ribbon, the QAH edge states are located at opposite edges of the ribbon
and the QSH edge states are restricted in a narrow region of a single edge.
In the presence of above three kinds of disorders, the conductance contributed by QSH edge states are easily to be suppressed and the QAH edge states remain intact.
The valley polarization is enhanced consequently. For moderate disorder strength, nearly fully valley polarized QAH edge states appear within the energy gap in which the edge states determine the transport. In above situations the intervalley scattering is insignificant.
These results are supported by the local current distribution as well.

The rest of the paper is organized as follows. The model of the valley filter, the methods and the parameters are detailed in section \ref{Model}. In section \ref{results}, we give the numerical results for the performance of the valley filter under three type of disorders (Anderson disorder, edge defects and edge deformations in sperate subsections), including the conductance, valley resolved transmission coefficients, valley polarization and local current distributions. A brief conclusion is given in section \ref{conclusion}.

\section{MODEL AND METHODS}\label{Model}

\begin{figure}
\includegraphics[width=\columnwidth, viewport=66 8 800 575, clip]{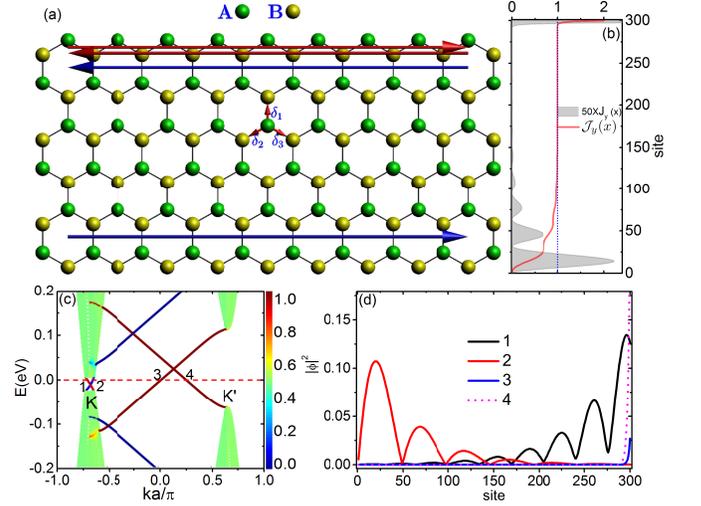}
\caption{(color online) (a) The schematic of $\rm{Sb}$ hexagonal lattice ribbon model in $\rm{Sb_2 H/LaFeO_3}$ heterostructure. The left and right sides are connected to two terminals of the same material with infinite length. Two sets of sublattices are represented by green (A) and yellow (B) balls. Three nearest neighbor vectors $\bm{\delta_1,\delta_2,\delta_3}$ are shown with red arrows. The edge states are indicated by transverse arrows at the zigzag boundaries: blue (red) arrows are for QAH (QSH) edge states. The width and length of the ribbon are $N=6$ (six atoms along the longitudinal direction) and $L=9$ (nine hexagonal lattice in the transverse direction), respectively.
(b) The spatial distribution and the cumulation of local current along the longitudinal direction.
Here the value of $J_y(x)$ is multiplied by $50$ and $\mathcal{J}_y(x) =
\sum_{x=1}^x J_y(x) $
is cumulated from the bottom upwards. The dotted blue line is $1.0$.
(c) The band structure of the model with the width $N=300$. The color represent the location of the position weight of the eigenstates. Two valleys are indicated by $K$ and $K'$, respectively.
The red dashed line is $E=0$. (d) Spatial distribution of wave functions along the longitudinal direction for eigenstates marked as $1-4$ in (c).
}\label{Fig1}
\end{figure}

The $\rm{Sb}$ atoms in $\rm{Sb_2 H/LaFeO_3}$ heterostructure formes
a hexagonal lattice as shown in Fig. \ref{Fig1} (a). $A$ and $B$ sublattices of
$\rm{Sb}$ atoms are coupled to $\rm{H}$ atoms and $\rm{Fe}$ atoms, respectively. We investigate a two terminal device composed of the central region and semi-infinite terminals. The disorders are considered in the central region.

In the tight-binding representation, the whole Hamiltonian ${H=H_1+H_2}$ can be divided into on-site term $H_1$ and the hopping term $H_2$,\cite{Q27}
\begin{eqnarray*}
H_1&=&\sum_{\bm{i}}\Phi_{\bm{i}}^\dagger[U_{\bm{i}} \tau_0\otimes\sigma_0+\lambda_{SO}\tau_z\otimes\sigma_z+M_{\bm{i}}\tau_0\otimes\sigma_z]\Phi_{\bm{i}},\\
H_2&=&\sum_{\bm{i}\in A}\sum_{j=1}^3 \Phi_{\bm{i}}^\dagger[T_{\bm{\delta_j}}+T_{R\bm{\delta_j}}]\Phi_{\bm{i}+\bm{\delta_j}}+H.c.
\end{eqnarray*}

Here the basis $\phi$ is $\phi=\{|\phi_+\rangle, |\phi_-\rangle\}\otimes\{ \uparrow,\downarrow\}$ with orbitals $\phi_{+}=-\frac{1}{\sqrt{2}}(p_x+ip_y)$, $\phi_{-}=\frac{1}{\sqrt{2}}(p_x-ip_y)$ and spin part $\{ \uparrow,\downarrow\}$. Thus, $\tau$ and $\sigma$ are the Pauli matrices acting in orbital and spin space, respectively. The first term in $H_1$ is the staggering potential with $U_{\bm{i}}=U(-U)$ in $A(B)$ sublattice. The second term in $H_1$ comes from the intrinsic spin-orbit coupling at site $\bm{i}$. The third term in $H_1$ is the staggering exchange field in A(B) sublattice contributed by $\rm{LaFeO_3}$ substrate and absorbed hydrogen atoms. Hamiltonian $H_2$ denotes the coupling between site $\bm{i}$ and its nearest sites $\bm{i+\delta_j}$ as the three vector $\bm{\delta_j}$ are displayed in Fig. \ref{Fig1}(a). The first and second term in $H_2$ represent the nearest hopping and the extrinsic Rashba spin-orbit coupling, respectively. The specific expressions of $T_{\bm{\delta_j}}$ and $T_{R\bm{\delta_j}}$ are
$T_{\bm{\delta_1}}={\left( \begin{array}{cc}t_1 & t_2 \\t_2 & t_1 \end{array} \right)}\otimes\sigma_0$,
$T_{\bm{\delta_2}}={\left( \begin{array}{cc}t_1 & z^2t_2 \\zt_2 & t_1 \end{array} \right)}\otimes\sigma_0$,
$T_{\bm{\delta_3}}={\left( \begin{array}{cc}t_1 & zt_2 \\z^2t_2 & t_1 \end{array} \right)}\otimes\sigma_0$ and
$T_{R\bm{\delta_1}}=-i{\left( \begin{array}{cc}\lambda_1 & \lambda_2 \\\lambda_2 & \lambda_1 \end{array} \right)}\otimes\sigma_y$,
$T_{R\bm{\delta_2}}=i{\left( \begin{array}{cc}\lambda_1 & z^2\lambda_2 \\z\lambda_2 & \lambda_1 \end{array} \right)}\otimes(\frac{\sqrt{3}}{2}\sigma_x
+\frac{1}{2}\sigma_y)$,
$T_{R\bm{\delta_3}}=i{\left( \begin{array}{cc}\lambda_1 & z\lambda_2 \\z^2\lambda_2 & \lambda_1 \end{array} \right)}\otimes(-\frac{\sqrt{3}}{2}
\sigma_x+\frac{1}{2}\sigma_y)$. Here $z=e^{2i\pi/3}$, $t_{1/2}$ and $\lambda_{1/2}$ are hoping term and Rashba spin-orbit coupling between nearest sites, respectively. The parameters adopted are $M_{A/B}=180/30meV$, $U_{A/B}=\mp25meV$, $\lambda_{SO}=220meV$, $t_{1/2}=1/-0.9eV$, $\lambda_{1/2}=10/-9meV$ for the case of QS-QAH state.\cite{Q27} Under such parameters, the band structure of the nanoribbon described by the Hamiltonian $H$ is shown in Fig. \ref{Fig1}(c) for the width $N=300$. The spatial location of each state is displayed with different colors. At the upper edge, there are three edge states and in the lower edge, there is only one edge state. The propagation of the edge states are also indicated in Fig. \ref{Fig1} (a): the blue (red) arrow indicates the QAH and the QSH state, respectively. Since only the $K$ states are well localized around $ka=-2\pi/3$, in the following discussion, the states other than $K$ are referred as $K'$ states which is ambiguous in the range of energy we are discussing: $[-0.04eV, 0.02eV]$.

The valley resolved transmission coefficients are calculated by the scattering matrix method. \cite{R24,A27,A28} The intervalley and intravalley components are summed up according to their valley indexes: $T_{K_1K_2}(E)=\sum_{k_1\in K_1}\sum_{k_2\in K_2}\tau_{k_1k_2}$ with $\tau_{k_1k_2}$ the transmission coefficient from state of $k_2$ (in momentum space) in left terminal to state of $k_1$ in the right terminal. Thus the transmission coefficient of electrons into valley $K$ in the right terminal is $T_K=T_{KK'}+T_{KK}$ (similarly $T_{K'}=T_{K'K'}+T_{K'K}$). In the numerical calculation, since the QSH edge states are not well located around $K'$ (see Fig. \ref{Fig1}(c)), we count states with $ka>-0.5\pi$ as valley $K'$ and $ka<-0.5\pi$ as valley $K$ for simplicity. The valley polarization of the current is $P=(T_K-T_{K'})/(T_K+T_{K'})$ and the linear conductance at zero temperature is $G=\frac{e^2}{h}(T_K+T_{K'})$. The method also applies to the newly reported ferrimagnetic honeycomb lattice for valley resolved transport calculation.\cite{Q25}

The local current distribution in two dimensional materials provides detailed information of carriers flow in space.\cite{A20,A26} To calculate it in our device, the non-equilibrium Green's function method is used.\cite{localcur1,localcur2,localcur3} The local current between
neighboring sites $\bm i$ and $\bm j$ is: $J_{\bm i\rightarrow \bm j}=\frac{2e^2}{h}Im[(T_{\delta}+T_{R\delta})(G^r\Gamma_LG^a)_{ji}](V_L-V_R)$ with $T_{\delta}$ and $T_{R\delta}$ the coupling between sites $\bm i$ and $\bm j$ and $V_L/V_R$ the voltage at the left and right terminal. Here $G^r/G^a$ is the retarded/advanced Green's function and $\Gamma_L$ the line-width function of the left terminal. In the numerical calculation, the correctness of local current distribution is verified that the summation $\sum_{x=1}^NJ_y(x)$ equals to $G(V_L-V_R)$ along every longitudinal line across the ribbon in the central region.

\section{Numerical results}\label{results}

First we investigate the spatial distribution of edge states in a clean zigzag edged ribbon. The band structure is shown in Fig. \ref{Fig1}(c).
Here and after $N=300$ is adopted, corresponding to a ribbon of width $450a$ with $a$ the lattice constant.
In Fig. \ref{Fig1}(c) the band gap at valley $K$ is much smaller than that of valley $K'$. In the QS-QAH regime (within the gap of valley $K$), there are four edge states.
The wave functions for the four cross points $1-4$ at $E=0$ are displayed in (d). For QAH states ($1$ and $2$), the wave functions show an oscillation behavior and mainly locate at two separated edges. The QSH states ($3$ and $4$), on the other hand, are strongly restricted at a single edge.
The different spatial distribution of wave functions comes from the fact that the gap in $K'$ is much larger than the gap at $K$. Furthermore, when there is a tiny voltage bias between the left terminal and the right terminal, the characters of local current $J_y(x)$ along the longitudinal direction is displayed in Fig. \ref{Fig1}(b). From bottom to top, $J_y(x)$ shows a periodical oscillation with decreasing peak value. On the other side, $J_y(x)$ is very high.
The cumulation of current, $\mathcal{J}_y(x) = \sum_{i=1}^x J_y(i)$, from bottom to site $x$ is also shown: it increases rapidly from $0$ monotonously and is saturated at $1.0$ in a wide range. At the other end, $\mathcal{J}_y(x)$ increases sharply and ends at $2.0$, corresponding to ballistic transport of $G=2e^2/h$.
It is clearly demonstrated again that the rightward QAH edge state is distributed at the lower boundary and the rightward QSH edge state is highly restricted at the upper boundary.
Besides, the two propagation modes belong to two separate valleys, leading to a current with no valley polarization.

In real experimental circumstances, two dimensional materials are usually accompanied with different types of disorders, such as Anderson disorder,\cite{Q10,A07} edge defects\cite{A21,A22} and edge deformations.\cite{A23,A24,A25} Thus one may imagine that if the material is not clean, the QSH edge states will be localized easily and leave the valley polarized QAH states along. So the device may perform as a perfect valley filter in the presence of disorders. Here Anderson disorder is introduced by the random on-site energy around the Fermi level. Edge defects mean atoms at the boundaries are randomly vacant with a certain probability. Edge deformations are corresponding to ribbons with curved edges. In the following, we will demonstrate them through numerical calculations.
\begin{figure}
\begin{center}
\includegraphics[width=\columnwidth, viewport=144 77 662 542, clip]{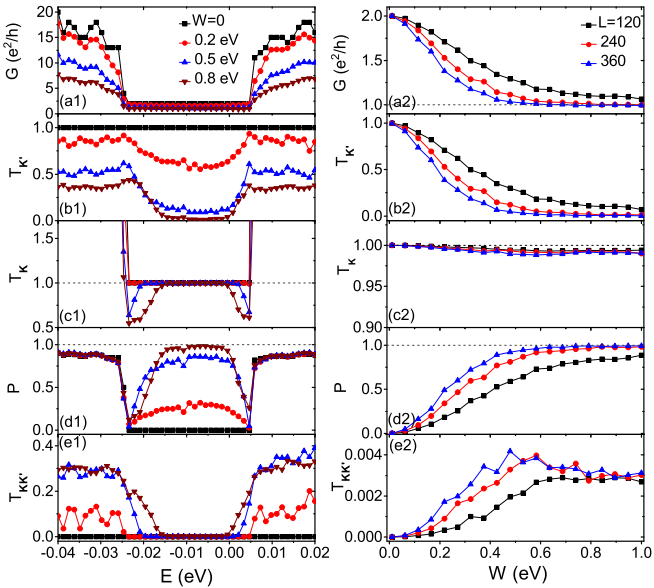}
\caption{(color online) The performance of the device ($G$, $T_{K/K'}$, $P$ and $T_{KK'}$) under Anderson disorder. (a1-e1) Quantities as function of the energy $E$ for different disorder strength $W$. (a2-e2) Quantities as function of $W$ for different disordered region length $L$. In the left column $L=240$ and in the right column $E=-0.01eV$. The dashed lines in (a2, c1, ,c2, d1, d2) indicate $1.0$.
}\label{Fig2}
\end{center}
\end{figure}
\subsection{Anderson Disorder}\label{results1}

When Anderson disorder is considered, an on-site random energy is added for each site. To do this, $H_W=\sum_{\bm{i}}\Phi_{\bm{i}}^\dagger\{ diag[\varepsilon_1,\varepsilon_2,\varepsilon_1,\varepsilon_2]\}\Phi_{\bm{i}}$ is added to $H$ with $\varepsilon_i$ the random potential uniformly distributed in the interval $[-W/2, W/2]$ and $W$ the disorder strength. For each $W$, $200$ disorder configurations are averaged. Here we focus on the energy region at $[-0.04eV, 0.02eV]$ because the gap center of valley $K$ locates around $E=-0.01eV$ and the gap is much smaller than that of valley $K'$ (see Fig. \ref{Fig1}(c)).

The results are displayed in Fig. \ref{Fig2}. In the clean limit, the conductance $G$ is large outside the gap and equals to $2e^2/h$ within the gap.
The quantized conductance is contributed equally by the QAH edge states and QSH edge states at opposite boundaries.
When the disorder $W$ is increased, $G$ is suppressed.
The transmission coefficients $T_K$ and $T_{K'}$ versus the energy $E$ is shown in (b1) and (c1) for different $W$.
For all $E$ values, $T_{K'}$ is suppressed for increasing $W$ and such behavior is more obvious within the gap region. $T_{K}$, on the other hand, is suppressed outside the gap region and are slightly affected within the gap.
For moderate disorder $W$ (e.g. $W=0.5$ and $0.8$), $T_{K'}$ is very small inside the gap. Meanwhile $T_K$ is almost unaffected around the gap center
(see $E\in [-0.015,0]$ for $W=0.8$ in Fig. \ref{Fig2}(c1))
and is affected only at the boundaries of the gap.
The great difference between the behaviors of $T_{K}$ and $T_{K'}$, also the main advantage of the present material, comes from the fact that QAH states are separate to each other while QSH states are located at the same edge of the ribbon.
Thus, the incoming electron along the QSH edge states are easily backscattered and the transport tends to be destroyed by disorder. The QAH edge states, on the other hand, are topologically protected, and it can only be backscattered into the opposite boundary, thus it is much more robust.

The intervalley scattering coefficients are shown in Fig. \ref{Fig2} (e1). Since two components of intervalley scattering coefficients, $T_{KK'}$ and $T_{K'K}$, are basically equal to each other under disorder average, only one of them is shown.
Three main characters are found: i) Generally $T_{KK'}$ is enhanced for increasing $W$; ii) $T_{KK'}$ is rather large outside the energy gap; iii) within the gap interval, $T_{KK'}$ is close to zero and is almost unaffected by Anderson disorder. Thus the valley index for carriers are well preserved within the gap, implying that the valley polarization $P$ could be high in the QS-QAH regime. Next, the valley polarization $P$ vs. $E$ relations are shown in Fig. \ref{Fig2} (d1). Outside the gap, $P$ is large since the $K$ valley dominates. In the QS-QAH regime, $P$ is zero for $W=0$ and $P$ increases as $W$ becomes large. For moderate $W$ (e.g. $W=0.8$), $P$ is nearly $1.0$ around the gap center, demonstrating a fully valley polarized current. At the band gap edges, $P$ shows two dips with value close to zero.

Next we investigate how the sample length affects the valley transport performance of the ribbon. For fixed $E$ ($E=-0.01eV$), the conductance $G$ decreases from $2.0$ to near $1.0$ as the disorder $W$ increases. For longer samples, $G$ decays much more rapidly for strong $W$. The behavior of $T_{K'}$ vs. $W$ is similar to that of $G$: it decrease from unity value to zero. For very large $W$, $T_{K'}$ is saturated with small magnitude (see Fig. \ref{Fig2}(e2)). The effect of disorder on $T_K$ is rather weak, even for large $W$ (e.g. $W=1$) and longer sample ($L=360$), $T_K\sim0.99$. Again, $T_{KK'}$ is rather small inside the energy gap. Finally, the valley polarization $P$ increases for increasing $W$ and $L$, it approaches unity value. So the transport of QAH edge state is dispassionless and of high valley polarization. The advantages are more obvious for larger $W$ and longer $L$. These behaviors mean that a perfect valley filter is achieved in the present device.

\begin{figure}
\includegraphics[width=\columnwidth, viewport=114 93 657 562, clip]{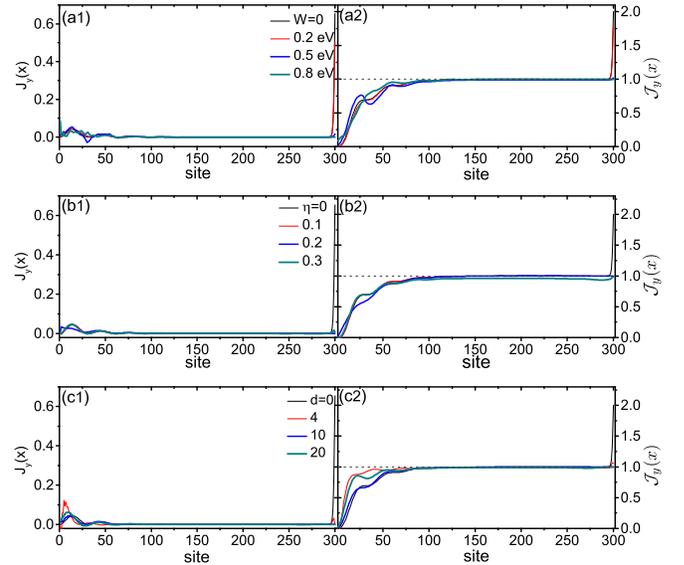}
\caption{(color online) The spatial distribution $J_y(x)$ (the left column) and the cumulation of local current $\mathcal{J}_y(x)$ (the right column) for Anderson disorder (a1-a2), edge defects (b1-b2) and edge deformations (c1-c2).  $L=240$ in all three cases.
}\label{Fig3}
\end{figure}

The local current distribution under Anderson disorder is displayed in Fig. \ref{Fig3}(a1) with $L=240$.
The curve of $J_y(x)$ for $W=0$ is a copy from Fig. \ref{Fig1}(b).
The translational invariance is broken under Anderson disorder,
so the mean value of $J_y(x)$ is taken at $20$ continuous sites attached to the right terminal of the scattering region. As the disorder is strengthened, $J_y(x)$ at the lower part shows non-periodical oscillatory behaviors and the value at the upper boundary is strongly suppressed.
For instance, for $W=0.5eV$, $J_x(300)$ is suppressed from $\sim0.6$ at $W=0$ to $\sim0.01$.
The cumulation of $J_y(x)$, $\mathcal{J}_y(x)$, is shown in Fig. \ref{Fig3}(a2).
At weak disorder $W$, $\mathcal{J}_y(x)$ shows no obvious different from that of an idea ribbon.
%On the right side of Fig. \ref{Fig3} (a2), $\mathcal{J}_y(x)$ is smaller than $2.0$.
For large $W$ (e.g. $W=0.5$), $\mathcal{J}_y(x)$ shows an oscillatory behavior
at the lower boundary of the ribbon and it saturates at $1.0$ at the upper boundary.
When $W$ is small, both $J_y(x)$ and $\mathcal{J}_y(x)$ are similar to an idea one in the lower half part of the ribbon and the value in the upper half part is suppressed.
At large $W$, $J_y(x)$ is almost destroyed at the upper boundary.
The result indicates that the QSH edge states are fragile to disorder even at weak disorder while the valley polarized QAH edge states are not affected.

\subsection{Edge Defects}\label{results2}
\begin{figure}
\includegraphics[width=\columnwidth, viewport=142 76 666 545, clip]{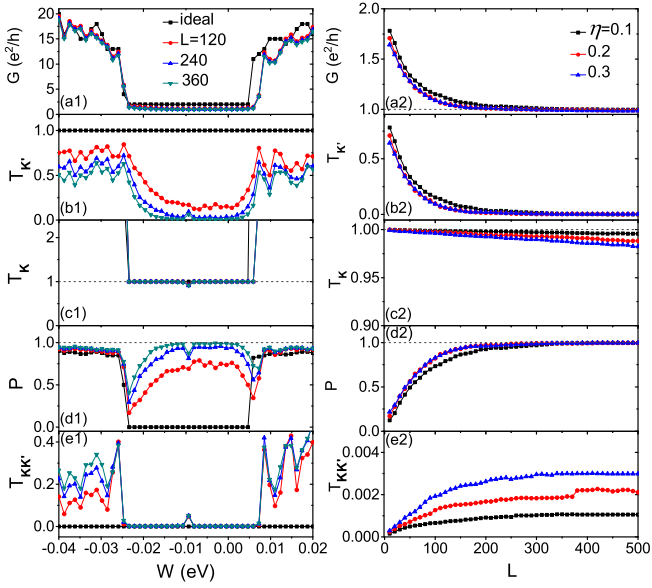}
\caption{(color online) The performance of the device ($G$, $T_{K/K'}$, $P$ and $T_{KK'}$) with edge defects. (a1-e1) Quantities as function of $E$ for different $L$. (a2-e2) Quantities as function of $L$ for different defect density $\eta$. In the left column $\eta=0.1$ and in the right column $E=-0.01eV$. The dashed lines in (a2, c1, c2, d1, d2) indicate $1.0$.
}\label{Fig4}
\end{figure}
In case of edge defects, the outmost $\rm{Sb}$ atoms at the edges are randomly removed with the probability $\eta$. In Fig. \ref{Fig4} (a1), the conductance $G$ vs. $E$ curves are shown for different length $L$ with $\eta=0.1$.
Within the gap, $G$ shows a plateau with the value close to $e^2/h$ for different $L$.
Meanwhile, the transmission coefficient $T_{K'}$ is severely suppressed but $T_{K}$ is almost unaffected as plotted by Fig. \ref{Fig4} (b1-c1).
The suppression is more obvious for large $L$, indicating stronger backscattering of QSH edge states for longer samples.
The results for intervalley scattering are shown in Fig. \ref{Fig4} (e1). Outside the QS-QAH regime, the intervalley scattering is strong. In contrast, within the QS-QAH regime, the intervalley scattering is almost zero. Because of the robustness of QAH edge state and the fragility of the QSH edge state, it is not surprise to see high value of valley polarization $P$ within the gap.
$P=1$ plateau is achieved when the length $L$ is large.
\begin{figure}
\includegraphics[width=\columnwidth, viewport=142 12 668 565, clip]{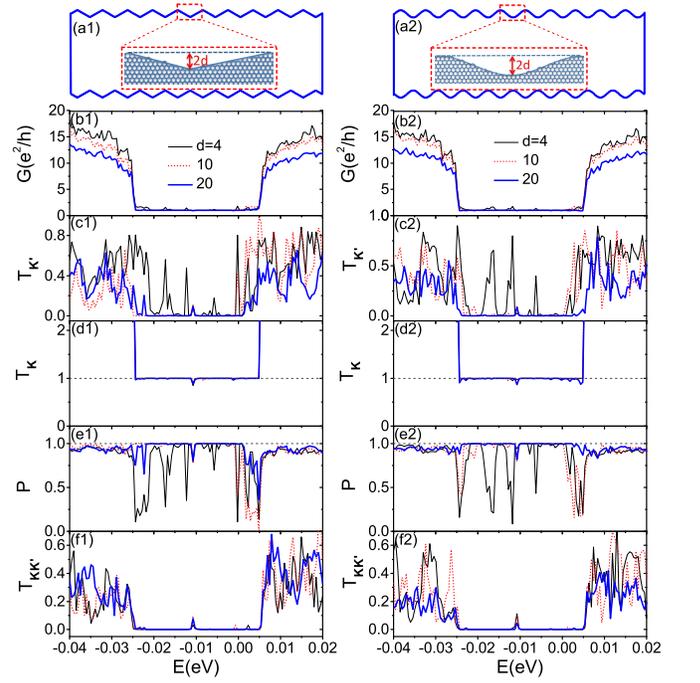}
\caption{(color online) The schematics of the sawtooth edged (a1) and sinuous edged (a2) ribbons with $10$ periods. The details of the edges are enlarged and in both cases the deformation is indicated by the maximum deviation from the original lateral edge, $d$.
The performance of $G$, $T_{K/K'}$, $P$ and $T_{KK'}$ as a function of $E$ for different types of edge deformations. The results for ribbons of sawtooth/sinuous edges are shown in (b1-f1)/(b2-f2).
}\label{Fig5}
\end{figure}

In Fig. \ref{Fig4} (a2-e2) we investigate the length dependence on the valley filter performance for different $\eta$ values.
the conductance $G$ decreases monotonically from $2e^2/h$ to $e^2/h$ as $L$ increases.
For large $\eta$ values, $G$ decreases much more rapidly.
For example, for $\eta=0.2$, $G$ is close to $e^2/h$ when $L>200$.
As expected, the decreasing behavior is mainly contributed by $T_{K'}$: as $L$ becomes large, $T_{K'}$ decay to zero rapidly while $T_K$ are only slightly affected.
Even for $\eta=0.3$ and $L=500$, $T_K$ is larger than $0.98$. The valley polarization $P$ vs. the length $L$ relations increase and are saturated with value $P=1$.
The higher value of defect density $\eta$, the faster $P$ becomes saturated.
For all $\eta$ values, the valley polarization $P$ reaches $1$ when $L>200$.
Again, the intervalley scattering for all cases (different values of $\eta$ and $L$) is rather weak (see Fig. \ref{Fig2} e2).

The local current distribution $J_y(x)$ for different $\eta$ is shown in Fig. \ref{Fig3} (b1) with $L=240$. The clean model result is also presented for comparison. For different $\eta$ values, the curves for $J_y(x)$ coincide each other in the most sites of the ribbon (except the upper boundary).
At the upper boundary, $J_y(x)$ decreases quickly from $\sim0.6$ to nearly zeros as $\eta$ increases.
The cumulation of current $\mathcal{J}_y(x)$ is shown as well in Fig. \ref{Fig3} (b2).
Similarly, in the most sites of the ribbon, the curves are only slightly changed.
As $\eta$ increases, the final value of $\mathcal{J}_y(x)$ decreases from $2.0$ to $1.0$.
It means only the QSH edge states are destroyed and the QAH edge states are nearly unaffected.
We note that the transport of QSH edge states vanishes even for small $\eta$, and the conductance $G=1$
and the valley polarization $P=1$ hold for even large $\eta$ values.
So in the presence of edge defects, the device can work as a perfect valley filter.

\subsection{Edge Deformations}\label{results3}
Finally we investigate the effect of the edge deformations on the performance of the valley filter.
Two categories of edge deformations are considered: the sawtooth edge (Fig. \ref{Fig5} a1) and sinuous edge (Fig. \ref{Fig5} a2). The size of the central region is $N=300$ and $L=360$. For both categories, at the edges there are $10$ periods. The deformation of the edge is indicated by $d$.
It means the maximum deviation of edge atom from an ideal ribbon (no deformation) is $da$ with $a$ the lattice constant.

The numerical results are shown in Fig. \ref{Fig5} with sawtooth edge in the left column and sinuous edge in the right column. For the first category of samples (left column), outside the gap region, the conductance $G$ is suppressed and inside the gap, $G$ vibrates above the value $G=e^2/h$.
From Fig. \ref{Fig5}(d1), one notices that for all $d$ values $T_{K}$ stays at $1.0$ in the QS-QAH regime, almost unaffected by the deformed edges.
However, $T_{K'}$ is strongly weakened except several peaks.
As the deformation is severe (larger $d$), the peaks inside the gap are suppressed and $T_{K'}$ tends to be flat with value zero.
Consequently, for ribbon with edge slightly curved (e.g. $d=4$), the valley polarization $P$ can reach the value of $1.0$ except several dips. For large $d$, $P$ equals to $1.0$ in almost whole QS-QAH regime. Similar to the cases for Anderson disorder and edge defects, $T_{KK'}$ vibrates outside the gap and is almost zero inside the gap under edge deformation.
For the second category of samples, the results (see Fig. \ref{Fig5} b2-f2) are quite similar.

Finally, we show current distribution results in Fig. \ref{Fig3} (c1-c2) for edge deformations.
In analogy to the situations for Anderson disorder and edge defects, when the ribbon edges are only slightly curved (e.g. $d=4$), at the upper boundary of the ribbon $J_y(x)$ is nearly zero and $\mathcal{J}_y(x)$ is $1.0$. These characters holds as $d$ increases from $4$ to $20$.
So in a ribbon of curved edges, QSH edge states are localized and the valley polarized QAH edge states stay robust. In a word, edge deformation, naturally exists in the fabricated process, facilitates the performance of $\rm{Sb}$ monolayer as perfect valley filters.

\section{Discussion and Conclusion}\label{conclusion}

The valley transport properties of zigzag edged monolayer $\rm{Sb}$ ribbon in $\rm{Sb_2 H/LaFeO_3}$ heterostructure is investigated.
Around the Fermi level, the system belongs to QS-QAH regime.
In such case, the QAH edge states are distributed at both sides of the ribbon and the states belong to the same valley. The QSH edge states, at the other valley, are located in a narrow region at the same side of the ribbon.
We find that under Anderson disorder, edge defects or edge deformations, the QSH edge states are easily to be localized and the valley polarized QAH edge states stay robust.
Consequently the current through the ribbon carried by quantized QAH state is of perfect valley polarization. The above results are both supported by transport calculation and local current distributions. So a $\rm{Sb}$ monolayer under disorders is a promising candidate for realizing perfect valley filters.

\section*{ACKNOWLEDGMENTS}

We thank Shenyuan Yang for helpful discussion.  H. Jiang thanks Y. J. Wu for careful reading the manuscript. This work was supported by NSFC under Grants Nos. 11674264, 11374219, 11534001, and 11574007,
NSF of Jiangsu Province, China (Grant No. BK20160007),
National Key R and D Program of China (2017YFA0303301) and
NBRP of China (2015CB921102).

\end{document}